\documentstyle[prb,aps,epsfig]{revtex}

\begin{document}
\baselineskip=12pt

\draft
\flushbottom
\twocolumn[\hsize\textwidth\columnwidth\hsize
\csname@twocolumnfalse\endcsname
\title {
Transport and Localisation in the Presence of Strong Structural and 
Spin Disorder}

\author{Sanjeev Kumar$^1$ and Pinaki Majumdar$^2$}

\address{                                   
$^1$~Institute for Physics, Theoretical Physics III, Electronic
Correlations and Magnetism,\\
  University of Augsburg, 86135 Augsburg, Germany.\\
$^2$~ Harish-Chandra  Research Institute,
 Chhatnag Road, Jhusi, Allahabad 211 019, India }

\date{May 22, 2005}

\maketitle
\tightenlines
\widetext
\advance\leftskip by 57pt
\advance\rightskip by 57pt
\begin{abstract}

We study a tight binding model including both on site disorder and  coupling
of the electrons to randomly oriented magnetic moments. The transport properties
are calculated via  the Kubo-Greenwood scheme, using the exact eigenstates of
the disordered system and large system size extrapolation of the low frequency 
optical conductivity.  We first benchmark our method in the model with only 
structural disorder and then use it to map out the transport regimes and metal-
insulator transitions in problems involving $(i)$~scattering from random magnetic 
moments, and $(ii)$~the combined effect of structural disorder and magnetic 
scattering.  We completely map out the dependence of the d.c conductivity on 
electron density $(n)$ the structural disorder $(\Delta)$ and the magnetic 
coupling $(J')$, and locate the insulator-metal phase boundary in the space of 
$n-\Delta-J'$. These results serve as a reference for understanding transport 
in systems ranging  from  magnetic semiconductors to double exchange `colossal 
magnetoresistance' systems. A brief version of this study appears in our earlier 
paper Europhys. Lett. {\bf 65}, 75 (2004).

\

\

\end{abstract}

]

\narrowtext
\tightenlines

\section{Introduction}

The most commonly studied case of localisation pertains to
non interacting electrons in the background of structural disorder.
There is a large body
of work \cite{lee-ram,kram-mck,mott-davis,mott-mit},
 analytical and numerical, as well as experimental studies,
that have focused on this problem. 
The principal qualitative result of these investigations 
is that in one and two dimensions 
all electronic eigenstates are localised for 
arbitrarily weak disorder, while in three dimension 
we need a critical disorder for complete localisation.  
In three dimension, at a given
disorder, all states {\it beyond}
an energy $\epsilon_c$ of the band center are localised and 
the system is metallic or insulating depending on whether the
Fermi level, $\epsilon_F$, lies in the region of extended states
or 
localised states. The `mobility edge', $\epsilon_c$, collapses to the band
center as the disorder is increased, driving the Anderson
metal-insulator transition (MIT). 

The presence of magnetic moments in a metal brings in several new
effects, depending on the strength of electron-spin coupling $(J')$, 
the concentration of moments $(n_{mag})$, the extent of disorder,
and the `character' (small or large $S$) of the moment.

In the `quantum limit', $2S \sim 1$, 
and for antiferromagnetic coupling,
the basic physics is contained in 
the Kondo effect.
For $n_{mag} \ll 1$,  the magnetic moments act as
`Kondo impurities' whose effects \cite{hewson}
 are now well 
understood.
For $n_{mag} \sim 1$, {\it i.e}, 
the concentrated Kondo limit, 
there can be various phases 
depending on 
electron-spin coupling and disorder. The ground state could be a
 non magnetic `heavy Fermi liquid' \cite{hf-ref1}, 
or a spin glass\cite{hf-ref2}, or a 
magnetically ordered state \cite{hf-ref3}. 
The physics of these Kondo lattice,
with quantum spins, is a vast area of research.
In this paper, however, we will avoid the issues of heavy fermion physics and
focus instead on electron-spin systems involving ``large $S$'', {\it i.e},
effectively `classical' moments.

For classical moments also, the effects vary depending on
$n_{mag}$, electron density, 
$J'$, and the extent of disorder. 
A wide variety  of magnetic systems
\cite{dms1,dms2,dms3,eub6-1,eub6-2,cmr-ref,gdsi1,gdsi2,f-mag-ref}
 are described, 
to a first approximation,
by electrons locally coupled to $d$ or $f$ moments, with $2S \gg 1$,
and moving in a 
structurally disordered background.
The magnetic ground state could be ferromagnetic, 
or a
more complicated ordered state, 
 or a spin glass. Transport often involves insulator-metal
transitions and colossal magnetoresistance.
The simplest Hamiltonian capturing these effects is:
\begin{equation}
H = -t \sum_{\langle ij \rangle, \sigma }   
c^{\dagger}_{i \sigma}c_{j \sigma}  
+ \sum_{i\sigma} (\epsilon_i - \mu) n_{i \sigma} 
- J'\sum_{\nu}  {\bf \sigma}_{\nu}.{\bf S}_{\nu}
\end{equation}

The $t$ are nearest neighbour hopping on a simple cubic lattice.
The random on site potential, 
$\epsilon_i$, is uniformly distributed between $\pm \Delta/2$.
The sites ${\bf R}_{\nu}$ are a subset of the cubic lattice
sites,
${\bf R}_i$, and correspond to the magnetic `dopant' locations.
Even with this simple model there are four 
dimensionless parameters in the problem. These are 
disorder $\Delta/t$, magnetic coupling $J'S/t$,
electron density $n$ (controlled by $\mu$), and the `density'
of moments $n_{mag}$. We will eventually study the $n_{mag} =1$ case,
but retain a more general structure right now. 
We absorb $S$ in our magnetic coupling $J'$, assuming  
$\vert {\bf S}_i \vert = 1$. 

Real materials have  band degeneracy and additional interactions
but the basic 
physics of several currently interesting materials
arise as limiting cases of the model above. 
$(i)$~The 
II-VI diluted magnetic semiconductor \cite{dms1,dms2,dms3}
 (DMS) 
Ga$_{1-x}$Mn$_x$As, 
exhibiting
high ferromagnetic $T_c$, correspond to 
$n_{mag} \ll 1$, $J'/t \sim 1$, weak disorder, and 
low electron density,
$n < n_{mag}$. 
$(ii)$~The Eu based magnetic 
semiconductors \cite{eub6-1,eub6-2}, EuB$_6$ etc,
involve $n_{mag} = 1$, since every Eu atom has a moment, $J'/t \gg 1$,
low carrier density, and possibly weak disorder. 
$(iii)$~The `colossal 
magnetoresistance' (CMR) manganites \cite{cmr-ref},
 specifically 
 La$_{1-x}$Sr$_x$MnO$_3$, involve
 $n_{mag} = 1$, $J'/t \gg 1$,
high electron density, and moderate `effective disorder'. To 
describe the 
more strongly resistive manganites,
the Ca doped systems, say,
 one requires additional 
electron-phonon interactions. 
$(iv)$~The amorphous magnetic
semiconductor \cite{gdsi1,gdsi2},
$a$-Gd$_x$Si$_{1-x}$, corresponds to $J'/t \gg 1$, $\Delta/t \gg 1$,
and $n_{mag} \sim n \sim {\cal O}(0.1)$.
Finally, $(v)$~the  traditional metallic $f$ electron magnets 
\cite{f-mag-ref}, correspond to
$n_{mag} =1$, and moderate to strong $J'$.

The  focus in the materials above is often 
on magnetism rather than localisation effects.
However, many of them have rather large resistivity in the 
paramagnetic phase, and $a$-GdSi, for example,
 shows a metal-insulator transition
at $T=0$ itself, on lowering carrier density.
Since 
there is no direct  spin-spin interaction in these systems, 
{\it the local electron-spin coupling controls both the
magnetic properties and the  character
of the electronic state.}

The intimate coupling between charge transport, localisation effects,
and magnetism in these systems  suggest
that we need to look beyond the traditional boundaries separating
`magnetism' from transport and localisation studies.
A complete study of electronic resistivity as a function of 
temperature, for any of the materials above, requires a solution
of the magnetic problem first. Since the moments are assumed
to be  classical, the electrons can be imagined to move in a
{\it static} background comprising the (quenched) structural
disorder and {\it annealed} spin disorder.
Evaluating the distribution of the annealed disorder is a 
non trivial problem, particularly in the strong coupling
(large $J'$) context that is experimentally relevant. We will
touch upon this in the next section, but this paper 
is concerned with transport 
and localisation effects in the
fully spin disordered phase.
In this limit, we  will 
present a comprehensive discussion of the resistivity
arising from the interplay of structural disorder and `paramagnetic'
scattering, and map out the metal-insulator phase diagram in terms
of electron density, disorder and magnetic coupling.

There have been some studies of electronic transport
in the background of random spins and structural disorder, 
acting independently or together.
Among these, the Anderson localisation problem
itself has
been extensively studied,
via  perturbation
theory \cite{and-loc1}, self-consistent schemes
\cite{and-loc2}, numerical techniques \cite{and-loc3}, 
and mapping to a 
field theory \cite{and-loc4}. 
Most of the qualitative issues in this context
are essentially settled.
{\it Weak}
magnetic scattering in a structurally  disordered system
has been studied \cite{and-loc-sf1,and-loc-sf2}
in the early days of weak localisation (WL)
theory to clarify the `dephasing' effect
of electron spin flip on quantum interference.
In the opposite 
limit of strong coupling, corresponding to double
exchange, localisation
effects have  been studied \cite{dex-loc} 
considering both magnetic and structural~disorder.

These efforts still 
leave a large and interesting part of $\Delta - J'$ space
unexplored.
To give a few examples, 
there is no discussion of the following:
$(i)$~the  resistivity
from purely magnetic scattering, as $J'$ 
rises through the perturbative regime to  double exchange:
this is the classic problem of paramagnetic scattering in 
`clean' magnets, studied earlier at weak coupling 
\cite{dg-friedel,fish-lang}.
$(ii)$~the effect of spin disorder on the 
Anderson transition, {\it i.e}, how the `anti-localising' effect
of spin flip scattering, at weak disorder, 
evolves into an insulator-metal transition (IMT). 
This is 
an instance of Anderson transition with broken 
time reversal symmetry, and
$(iii)$~the wide `middle', where the effect 
of neither $\Delta$ nor $J'$ is perturbative 
and their contribution to the  resistivity is not additive 
({\it i.e}, violates Mathiessens rule).
This is the regime relevant to DMS,  CMR materials,
and amorphous magnetic~semiconductors.

The next section describes the transport calculation in
detail. Following that we present results on transport, 
successively, in the structural disorder problem,
the magnetic disorder problem, and the simultaneous effect
of both. 
This paper follows up on our earlier short paper \cite{sk-pm-epl}.

\section{Computational scheme}

Although we will work with random uncorrelated
spins, viewing the magnetic disorder as quenched,
 let us highlight
how the `true' spin distribution can be evaluated, and the limit
where the background can be considered random.
Following that we describe our transport calculation method.

\subsection{The spin distribution}

The `structural' variables $\epsilon_i$ are quenched, and have a 
specified distribution, but 
the spin orientations ${\bf S}_i$ are not known
{\it a priori}. The system chooses a spin configuration, at
$T=0$, to optimise the total energy. To calculate the `true'
ground state properties, or finite temperature transport,
 we need to solve for the spin distribution first and then 
evaluate electronic properties in these spin background.
Denoting the full spin configuration as 
$ \{ {\bf S}_i \} $, the  spin distribution
$P\{ {\bf S}_i \} $  is given by:
\begin{eqnarray}
P \{ {\bf S}_i \}  &=&    Z^{-1} Tre^{-\beta H}  \cr
Z &=& \int {\cal D}{\bf S}_i Tr e^{- \beta H} 
\nonumber
\end{eqnarray}
where $Z$ is the full partition function of the system, and the
`trace' is over fermionic variables. 
Equivalently, the
effective classical `Hamiltonian' controlling the Boltzmann
weight for spins is: 
\begin{equation}
H_{eff}\{ {\bf S}_i \} = 
-{1 \over {\beta}}  log~Tr e^{- \beta H} 
\end{equation}
$H_{eff}$ is the 
fermion (free) energy {\it in the background }
$\{ {\bf S}_i \} $.

To make more sense of the formal expression above, consider
 $J'/t \ll 1$. In this case we can expand  the fermion (free)
energy 
in powers of $J'$. For a non disordered system this leads to
the standard RKKY coupling \cite{rkky-ref}
 between the classical spins, while
the 
presence of structural
disorder, leads to a `bond disordered' RKKY model:
$
H_{eff} \sim \sum_{ij} J_{ij} {\bf S}_i . {\bf S}_j,
$
where the exchange 
$J_{ij}$ are $ \sim J'^2 \chi_{ij}$ the  
 $\chi_{ij}$ being the 
non local spin response function of the disordered,   
 $J' = 0$,
electron system.
Having obtained the effective spin Hamiltonian, the 
transport properties are to be calculated 
by considering electron motion
in the backgrounds $\{ \epsilon_i , {\bf S}_i \}$ where the $\{ {\bf S}_i \}$
are equilibrium configurations of $H_{eff}$ for a specified
realisation of disorder $\{ \epsilon_i \}$.

At strong coupling, {\it i.e}, 
large $J'$, the fermion trace cannot be 
analytically evaluated, and it is 
no longer possible to write an explicit  spin Hamiltonian.
We need special techniques to anneal the spins.
The magnetic order and the complete transport properties in 
such (disordered) Kondo lattice models
is discussed elsewhere \cite{scr,dde}.

The complications of the magnetic problem can be avoided if
we {\it assume} a spin distribution. 
The simplest distribution one can assume corresponds to
{\it uncorrelated random spins}. This is physically relevant in
two limits.
 
$(i)$~At sufficiently high temperature, compared to the magnetic 
ordering scales in the problem, the spins are essentially
randomly fluctuating, with only short range correlation. 
The magnetic 
ordering scale for $J'/t \ll 1$ is $\sim f_1(n) J'^2/t$, while
for $J'/t \gg 1$ the ordering scale is $\sim  f_2(n) t$, where 
$f_1$ and $f_2$ are electron 
density dependent dimensionless 
coefficients and $f_2^{max} \sim 0.1$.
Compared to the typical Fermi energy, $\sim zt$, where $z$ is the
coordination number of the lattice, these scales are all small.
We use a $T=0$ formulation for transport, {\it i.e}, we do
not use Fermi factors,  but given 
the smallness of $T_c/\epsilon_F$, our results would be relevant
even in the `real' paramagnetic phase.
$(ii)$~Another situation in which a random spin configuration is
appropriate is a `spin glass', likely to occur in strongly
disordered systems \cite{sg-disord-ref}. 
There are always short range correlations
in a spin glass but if we ignore their effect on transport then
at all temperature the transport in such a system can be described,
approximately, in terms of a random spin background.

\subsection{Conductivity calculation}

In the linear response regime, the  Kubo formula
can be used to calculate the conductivity of a system. The
general expression \cite{mahan}, 
involving matrix elements between
many body states, simplifies significantly for 
non-interacting systems.
This `Kubo-Greenwood' result can be computed purely in terms of
single particle eigenfunctions and energies.

The numerical difficulty with this result lies in implementing
it for a finite size system, where the spectrum is discrete,
with gaps ${\cal O}(1/N)$, with  $N$ being the number of sites in the
system. Since the `d.c' conductivity 
involves transitions between essentially 
degenerate states at $\epsilon_F$, 
it cannot be 
calculated with control on small systems. As a result, instead
of computing the `Kubo conductivity'
it is more usual to compute the `Landauer conductance', $G$, of a 
finite system coupled to leads \cite{dutta}, 
and infer the conductivity from
the length dependence of $G$. 

We pursue the Kubo approach, indirectly, through a calculation
of the low frequency optical conductivity 
for a $L_T \times L_T \times L$ geometry.
The advantage 
of calculating the conductivity this
way is,  $(i)$~it ties in with diagonalisation that one may have to
do for the magnetic problem, and 
$(ii)$~it allows access to the optical
conductivity, without added effort, and can reveal the 
significantly non Drude nature of $\sigma(\omega)$ at 
strong disorder.
The principal 
disadvantage is, this scheme  
cannot be pushed 
beyond $N \sim 10^{3} - 10^4$,
and is therefore not useful for accessing critical properties.

For disordered non interacting systems, the Kubo formula, at $T=0$, is:
\begin{equation}
\sigma ( \omega)
= {A \over N}
\sum_{\alpha, \beta} (n_{\alpha} - n_{\beta})
{ {\vert f_{\alpha \beta} \vert^2} \over {\epsilon_{\beta} 
- \epsilon_{\alpha}}}
\delta(\omega - (\epsilon_{\beta} - \epsilon_{\alpha}))
\end{equation}
with $A = {\pi  e^2 }/{{\hbar a_0}}$, $a_0$ being the lattice spacing,
and $n_{\alpha}= \theta(\mu
- \epsilon_{\alpha})$. 
The $f_{\alpha \beta}$ are matrix elements of the current operator
 $j_x = i t  \sum_{i, \sigma} (c^{\dagger}_{{i + x a_0},\sigma}
c_{i, \sigma} - h.c)$, between exact single particle eigenstates
$\vert \psi_{\alpha}\rangle$, 
$\vert \psi_{\beta}\rangle$, {\it  etc},  and
$\epsilon_{\alpha}$, $\epsilon_{\beta}$ 
are the corresponding eigenvalues.

The conductivity above is prior to disorder averaging.
Notice that the $\delta$ function constraint cannot 
be satisfied for arbitrary frequency
in a finite system.
So we can neither calculate the d.c conductivity, $\sigma_{dc}$,
directly, nor estimate $\sigma(\omega)$ at some arbitrary externally
specified frequency.
However, we can still calculate the `average'
conductivity over a frequency interval $\Delta \omega$, defined below, 
and we use the following strategy to extract $\sigma_{dc}$.

$(i)$~The average of $\sigma(\omega)$  over the 
interval $[0, \Delta \omega]$ is defined as
\begin{equation}
\sigma_{av}(\Delta \omega, \mu,  N)
= {1 \over {\Delta \omega}}\int_0^{\Delta \omega}
\sigma(\omega, \mu, N)d \omega
\end{equation}
$\Delta \omega$ can be set independent of $N$,  but we
will relate them by fixing:
$\Delta \omega = B/N$.
We fix $B$ by setting  $\Delta \omega = 0.04$ 
for $N =1000$. The mean finite size gap is $12/1000 \sim 0.01$,
in 3d,
much smaller than $\Delta \omega$.

$(ii)$~$\sigma_{av}$ is averaged over $N_{r}$ realisations of
disorder, to generate ${\bar \sigma}_{av}(\Delta \omega, \mu, L)$.
The `noise' in  ${\bar \sigma}_{av}(\Delta \omega, \mu, L)$ falls slowly,
as 
$1/\sqrt{N_r} $.  We use $N_r \sim 100$ for the largest
sizes, to keep the computation  reasonable, and use a filter to
smooth the data over a small window in $\mu$. 

$(iii)$~We study the ${\bar \sigma}_{av}(\Delta \omega, \mu, L)$ for
$L_T=6$ and 
the sequence $L=24$ to $L=64$ in increments of $8$
and 
extrapolate, $ 
\sigma_{calc}(\mu) = lim_{L \rightarrow \infty}
{\bar \sigma}_{av}(\Delta \omega, \mu,  L)$.
As specified before, $\Delta \omega = B/N$. 

To calculate the full,  disorder averaged, optical 
conductivity we use the inversion:  
$ 
\sigma(\omega) = {\bar \sigma}_{av} (\omega) + \omega 
{ {d { \bar \sigma_{av}}} \over { d \omega}}$.
The $\sigma(\omega)$ results in this paper 
are mostly based on a  
$6 \times 6 \times 32$ geometry.

\section{Transport in the Anderson model}

The metal-insulator phase boundary and the critical properties near the 
transition have been extensively 
studied \cite{and-loc1,and-loc2,and-loc3,and-loc4}
 in the Anderson model. However,
 the actual resistivity seems to 
have received much less attention. 
As recently pointed out by Nikolic and Allen \cite{nik-allen}, 
there is a wide
regime in $\Delta$, between the Born-Boltzmann end and the 
scaling regime, where there are no analytic theories
of transport.
We study this `old problem' in some detail because the wealth of
existing results provides a benchmark for our method. There are 
very few exact results with which we will be able to
compare our data in the 
magnetic scattering problems.
 
\subsection{Global features}

The `global features' of transport and localisation in the Anderson
model are contained in Figs.1-3. The data is obtained via the
extrapolation procedure described earlier. Fig.1.(a) highlights the 
suppression in conductivity with increasing disorder, across
the entire band. The `weakest' disorder in this case, $\Delta =4$,
is probably already outside the semiclassical Boltzmann regime.
A naive $\Delta^2$ scaling of the resistivity still 
works, at the band
center, between $\Delta =4$ and $\Delta =8$, but the same extrapolated
to $\Delta = 16$, would imply $\rho(16)/\rho(4) \sim 16$, while 
the ratio  is actually $\sim 10^3$. 
This figure captures the expected 
crossover from moderate scattering, roughly following Boltzmann
scaling,  to localisation as $\Delta 
\rightarrow \Delta_c \sim 16.5$, the critical disorder \cite{slevin}
at the band center.
It also provides a glimpse of 
how the `mobility edge' moves with increasing disorder,
better quantified in Fig.3.
Note that for data 
at a specified system size, $L = 16, 32$ etc, shown later, 
the notion
of a `mobility edge' does not make sense, and all we observe is
a crossover
from small to large conductivity as $\mu$ is varied.
{\it The  change in $\sigma(\mu, L)$ with $L$, and the
$L \rightarrow \infty$ extrapolation, is 
crucial for identifying the mobility edge.}

The DOS plot, Fig.1.(b), illustrates the band broadening, and in Fig.3.(b)
we have compared our band edge energy with earlier CPA 
results \cite{econ-souk}.
The (algebraic) 
average DOS is featureless and non critical and does not
play an interesting role in the problem.

Since the band broadens significantly with disorder, 
$\sigma(\mu)$ by itself does not provide the {\it density}
dependence of the conductivity. Fig.2 takes into account 
the shift in $\mu$ required to maintain constant density (with 
increasing disorder) and shows $\sigma(n)$.
Our density is defined as average
number of electrons per site, so $n_{max} =2$. Since the 
model is particle-hole symmetric we show only the regime
$n = [0,1]$.

\begin{figure}
\vspace{.4cm}
\centerline{
\psfig{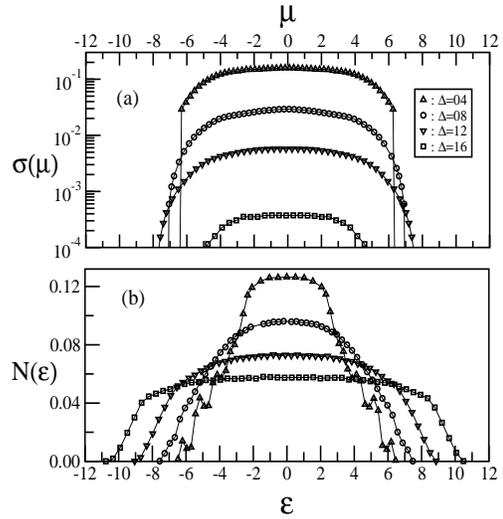}}
\vspace{.1cm}
\caption{Panel  $(a)$. Variation of conductivity with $\mu$,
and panel $(b)$. density of states, for  several values of~$\Delta$.}
\end{figure}

\begin{figure}
\vspace{.4cm}
\centerline{
\psfig{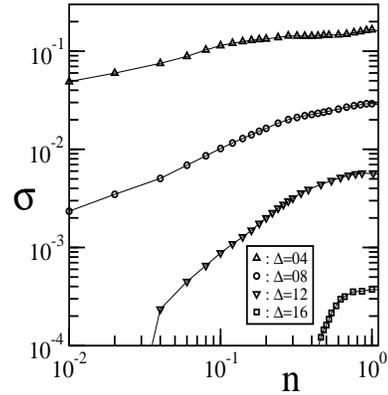}}
\vspace{.2cm}
\caption{ Variation of conductivity with  carrier density, for several
$\Delta$, 
 constructed from the $\sigma(\mu)$ and $N(\epsilon)$
data in Fig.1}
\end{figure}

\begin{figure}
\centerline{
\psfig{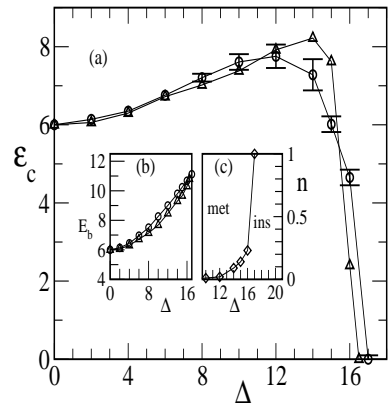}}
\vspace{.2cm}
\caption{ Main panel, $(a)$, shows the variation in 
mobility edge with disorder. We compare our results, circles,
with earlier work \cite{econ-souk}, triangles. Inset $(b)$ shows 
the `band edge', and $(c)$ the
fraction of localised states at large $\Delta$.
}
\end{figure}

To get a feel for the magnitude of the conductivity, which
we measure in units of $\pi e^2/({\hbar} a_0)$, note that
the  Mott `minimum metallic conductivity', $\sigma_{Mott}$,
at  the band center \cite{mott-mit}, 
is roughly $\sim 0.03 e^2/({\hbar} a_0)$.
Our dimensionless conductivity $\sigma_{calc}$, shown in the figures,
can be converted to
real units, $\sigma_{actual}$, by using
$$
\sigma_{actual} \sim 100*\sigma_{Mott}*\sigma_{calc}
$$ where 
we use $\sigma_{Mott} = 
 0.03 e^2/({\hbar} a_0)$.
The results we show 
in the present spin degenerate problem includes a factor of 
$2$ to account for the two spin channels.
This is important to 
compare with the magnetic scattering problems
later. The conductivity per spin channel
falls below $\approx 10^{-2}$ for 
$\Delta \gtrsim 8$. This implies that beyond
$\Delta \approx 8$, 
$\sigma < \sigma_{Mott}$
in the  Anderson model.

The main panel in Fig.3 shows the variation in mobility edge
with increasing disorder.
Our error estimates are based on the shift in $\epsilon_c$
as we change from moderate to zero filtering of the ${\bar \sigma}(\mu, L)$
data.
We show some earlier standard result \cite{econ-souk} 
for comparison. The best
current result on $\Delta_c$ is $16.5$, our method yields $\Delta_c 
\sim 17$. Our results on the band edge, Fig.3.(b), match reasonably
with earlier CPA based results.
Note that while the mobility edge has a `re-entrant' behaviour,
the fraction of localised states in the band, Fig.3.(c), increases
monotonically with disorder.

\subsection{Transport regimes}

There are tentatively three transport regimes in the Anderson
model. These are $(i)$ the perturbative Born scattering regime,
described by the Boltzmann transport equation and the low
order 
corrections in 
$(k_Fl)^{-1}$.
This corresponds to $\Delta/W \ll 1$, where $W = 12t$.
$(ii)$~The wide intermediate coupling regime $\Delta/W 
\sim {\cal O}(1)$, and  
$(iii)$~the `scaling' region, $\Delta \rightarrow \Delta_c$,
near the MIT.

\begin{figure}
\vspace{.2cm}
\centerline{
\psfig{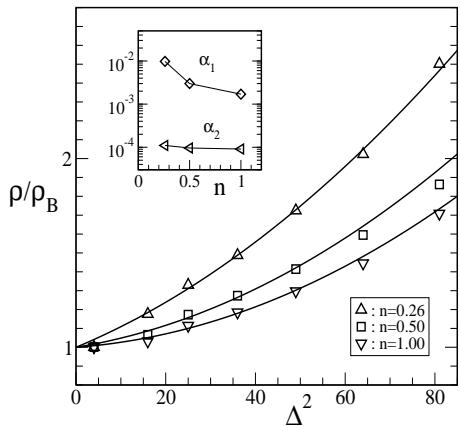}}
\vspace{.2cm}
\caption{ Variation of the resistivity, normalised to the
Born resistivity, with disorder. The firm lines are fits of the
form $1 + \alpha_1(n) \Delta^2 + \alpha_2(n) \Delta^4$. Inset shows
the coefficients $\alpha_1, \alpha_2$. } 
\end{figure}

We analyse the data, with increasing $\Delta$,
in the sequence $(i) \rightarrow (ii) \rightarrow (iii)$.

\subsubsection{Perturbative regime}

To leading order, the scattering rate from the disorder is
$\tau^{-1}_{\Delta}~\sim~2\pi N(\epsilon_F) 
\langle \epsilon_i^2 \rangle$. 
The second moment of the random potential is, 
$\langle \epsilon_i^2 \rangle \sim \Delta^2/12$. 
Since 
 $ N(\epsilon_F) \sim 0.13$, at the band center, Fig.1.(b), 
the 
scattering rate, $\Gamma_{\Delta} = \tau^{-1}_{\Delta}
\approx  \Delta^2/(15 t)$.

The three related quantities which define Boltzmann transport
are
$(i)$~the scattering rate, $\Gamma_{\Delta}$, defined above,
$(ii)$~the (inverse) mean free path $a_0/l \sim 0.03 (\Delta/t)^2$
at the band center, and $(iii)$~the Born-Boltzmann conductivity
$\sigma_B \approx 1.62(\pi e^2/{\hbar a_0})*(t/\Delta)^2$ per
spin channel, again at the band center.
In addition the optical conductivity should
have the Drude form $\sigma(\omega) = \sigma_B/( 1 + 
(\omega^2/\Gamma_{\Delta}^2))$,
and the `width' in the optical conductivity can be checked against
the magnitude of d.c conductivity.

Using the form for  $\sigma_B$, the conductivity at
$\Delta=4$, assuming Boltzmann transport, would be  
approximately $0.1 (\pi e^2/{\hbar a_0})$ per spin channel, {\it i.e}
$\sim 0.2$ in our units including spin degeneracy. Our data,
Fig.2, gives a value $\sim 0.19$ at the band center.
The crude Boltzmann 
 scaling is obvious from the moderate $\Delta$ results in
Fig.1 and Fig.2. In Fig.4 we attempt to quantify the corrections
to the Boltzmann result, still staying far from the localisation
regime.

The weak localisation  corrections that arise beyond
Boltzmann transport 
control the resistivity in one and 
two dimension. These are quantum
interference effects, formally arising from the `Cooperon' corrections.
A similar argument would lead us to believe that in
three dimension \cite{lee-ram}
the leading 
correction beyond the Boltzmann results should be $\delta \sigma
\propto  -(k_Fl)^{-1}$. Since $(k_Fl)^{-1} \propto  \Delta^2$ and the
Boltzmann conductivity $\sigma_B \propto  k_F l$, 
the net conductivity 
would be expected to have the form 
$\sigma \sim \sigma_B(1 - {\cal O}((k_Fl)^{-2}))$, {\it i.e},
$\sigma(\Delta) \sim \Delta^{-2} (1 - {\cal O}(\Delta^4))$.
In that case, the 
{\it resistivity} should have a form $\rho(\Delta)
\sim \rho_B(\Delta)(1 + {\cal O}(\Delta^4))$.

Fig.4 shows $\rho(\Delta)/\rho_B(\Delta)$ plotted against $\Delta^2$
for three densities. We avoid too low a density to keep the scales
comparable. The data are fitted to  $\rho/\rho_B = 1 +
\alpha_1(n) \Delta^2 + \alpha_2(n) \Delta^4$, upto $\Delta^2= 49$
and then extrapolated to $\Delta^2 = 81$. 

There are two notable features: 
$(i)$~There is clearly a
non zero coefficient $\alpha_1(n)$ so the equivalent of the WL
corrections do not control the leading correction to $\sigma_B$ in
three dimension.
The coefficients $\alpha_1$ and $\alpha_2$ are shown in the inset
in Fig.4.
$(ii)$~The `low $\Delta$' fit seems to
work reasonably for $\Delta \lesssim 8$, in the sense
that
$\rho/\rho_B \lesssim 2$. This {\it qualitative
correspondence}
 with the Boltzmann result, 
even in the regime $a_0/l \gtrsim 1$, 
has been noticed recently \cite{nik-allen}. 

The first issue has been explored in detail \cite{bel-kirk-3d}
 by Belitz and Kirkpatrick
who argue that the standard WL processes do not exhaust the leading
corrections to $\sigma_B$ in three dimension. 
According to them, the perturbative
expansion for $\sigma$, in a continuum model, has the form
$$
\sigma \sim \sigma_B  \{ 1 - a (k_Fl)^{-1} - b (k_F l)^{-2} log(k_F l) + 
{\cal O}((k_Fl)^{-2}) \} 
$$ 
where $a$ and $b$ are numerical coefficients ${\cal O}(1)$.
The WL argument would put $a=0$, $b=0$.

This form for the correction beyond Boltzmann
transport has apparently  been observed for electron mobility in dense
neutral gases. The detailed coefficients in this expression would
change in a tight binding model, but the key result about $k_Fl$
dependence should survive.

\subsubsection{Intermediate coupling}

The Boltzmann result alongwith the perturabtive quantum corrections is
reasonable probably upto $\Delta/W \sim 0.2-0.3$, although numerically
the fit, in the last section,
 seems to describe the resistivity even upto $\Delta/W \sim 0.75$.
The scaling regime, where localisation effects become visible,
 occurs  within about
$10 \%$ of $\Delta_c$.

Despite the correspondence of our numerical results with an
extrapolation of weak coupling theory, there is no analytic 
framework for calculating the resistivity when the ``small
parameter'' $(k_Fl)^{-1} \sim a_0/l$ becomes
${\cal O}(1)$. 

The  paramter $a_0/l$  is ${\cal O}(1)$ for 
$\Delta/W \sim 0.5$ but the deviation from the
Boltzmann result (at the band center) is only about $25 \%$. This 
has been pointed out recently by Nikolic and Allen \cite{nik-allen}
and probably arises from a fortuitious 
cancellation of higher order corrections.
The self-consistent theory (SCT) of Vollhardt and Wolfle \cite{and-loc2}
 serves
as an interpolating approximation in this regime. 
\begin{figure}
\vspace{1.0cm}
\centerline{
\psfig{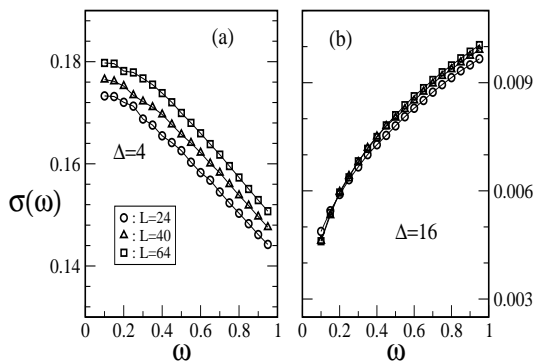}}
\vspace{.2cm}
\caption{ The optical conductivity, at band center, for different $L$
and $\Delta$. Panel $(a).$~corresponds to the moderate disorder regime,
with a Drude form for $\sigma(\omega)$, while $(b).$~is for a system 
on the verge of localisation (vanishing $\sigma_{dc}$).}
\end{figure}
\begin{figure}
\vspace{.3cm}
\centerline{
\psfig{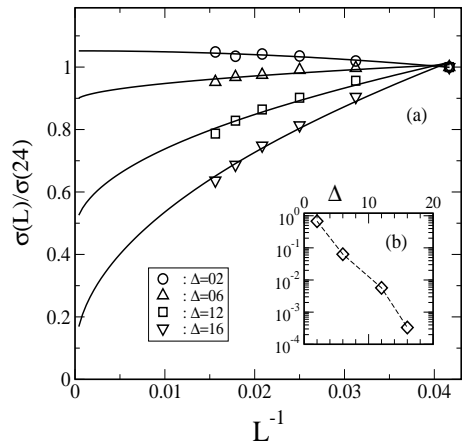}}
\vspace{.2cm}
\caption{ The approach to the d.c conductivity, with increasing $L$ in 
the $L_T \times L_T \times L$ geometry. $L_T =6$ 
and the disorder is increased from the perturbative end towards 
localisation.  The chemical potential $\mu = 0$, so $n=1$.
${\bar \sigma}(L)$, defined earlier in the text, is the average of 
$\sigma(\omega)$
over the interval $[0, \Delta \omega]$, with $\Delta \omega =
1.1/L$, at $L_T=6$.
Inset: variation of ${\bar \sigma}(L=24)$ with $\Delta$, to illustrate
the rapid fall in the reference 
conductivity with increasing disorder.}
\end{figure}

Within the SCT  
also, the conductivity at band center remains within
$20 \%$ of $\sigma_B$ for $\Delta \lesssim 8$.
The deviation from the Boltzmann result grows as we move
from the band center to the band edge as evident in Fig.4.

\subsubsection{Scaling regime}

The scaling regime occurs close to critical disorder, within about
$10 \%$ of $\Delta_c$. The conductivity in this regime varies
as $\Delta_c - \Delta$. This regime has been extensively studied
to clarify the critical properties (see, {\it e.g}, \cite{and-loc2}
and references therein). We have not used a dense 
enough sampling in $\Delta$ for discussing the critical behaviour,
and our 
system sizes too are not large enough for high accuracy 
calculation of the conductivity in this regime.  
However, based on results at $\Delta =16$ and $\Delta =17$
we can bracket the critical point, as shown in Fig.3.

\subsection{Optical conductivity}

The optical conductivity $\sigma(\omega)$ is of intrinsic interest
\cite{weisse}
and also plays a role in our method of determining the d.c 
conductivity.  There are some exact results known on the form
of the low frequency $\sigma(\omega)$ in the Anderson model.

$(i)$~At weak disorder, when Boltzmann transport holds, 
the optical
conductivity has the Drude form, $\sigma(\omega) \sim
\sigma(0)/( 1 + \omega^2 \tau^2)$, where $\tau^{-1}
\propto \Delta^2$ as we already know. For $\omega \tau \ll 1$
this would give us  $\sigma(\omega) \sim
\sigma(0)(1 - \omega^2 \tau^2)$. 
$(ii)$~When the quantum corrections to the d.c conductivity become
important the frequency dependence also picks up a non Drude
form. In the intermediate disorder regime, one expects
$\sigma(\omega) \sim \sigma(0) + 
{\cal O}(\Gamma \sqrt{\omega/\Gamma})$,
where $\sigma(0)$ already incorporates corrections beyond the
Boltzmann result. In this regime the conductivity {\it rises}
with increasing frequency, for frequencies $\omega \ll \Gamma$.
$(iii)$~At the
critical point, where the zero frequency conductivity vanishes,
$\sigma(\omega) \sim \omega^{1/3}$, and in the localised regime
$\sigma(\omega) \sim \omega^2$.

These results originally obtained through different techniques
can be obtained in a unified way via the self-consistent theory
of Vollhardt and Wolfle. 

Fig.5 demonstrates the changing character of $\sigma(\omega)$, at $n=1$, 
as we move from the Boltzmann regime $(\Delta =2)$, to strong 
disorder $(\Delta =16)$. 
We show the data for three system sizes at each $\Delta$ 
to illustrate the explicit $L$ dependence in $\sigma(\omega, L)$.
This is important for analysing the extrapolation needed for
$\sigma_{dc}$.

There are two effects of changing system size: $(i)$~the 
$\sigma(\omega)$ profile itself can change with evolving system size, 
over the
frequency range of interest, and $(ii)$~larger system size allows access to
(more dependable) low frequency data.

Fig.5.(a), the weak disorder case, reveals that 
the $\sigma(\omega)$ profile changes perceptibly with increasing $L$,
the changes being ${\cal O}(5 \%)$.
This implies that in our attempt to access d.c conductivity, the 
contribution arises not only from lowering the frequency cutoff
but also moderate changes in the  $\sigma(\omega)$
profile. At strong disorder,
Fig.5.(b), the profile itself does not change significantly with $L$
and the key change in the $\sigma_{dc}$ estimate comes from
our ability to access lower frequencies.

\subsection{Large $L$ extrapolation}

How important is the large $L$ extrapolation to access the
d.c conductivity, {\it i.e}, 
what is the error if we treat the 
\begin{figure}
\vspace{.7cm}
\centerline{
\psfig{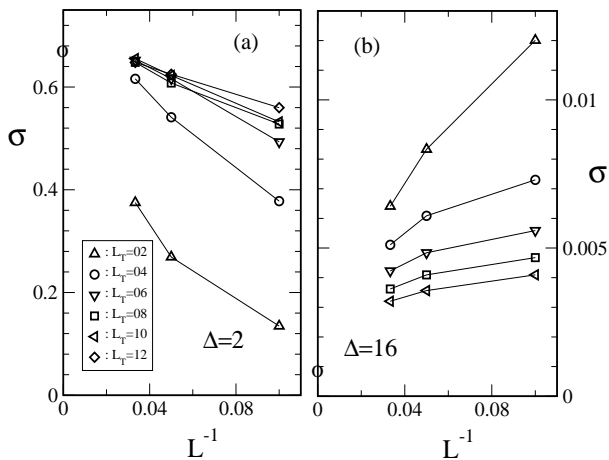}}
\vspace{.2cm}
\caption{ Effect of transverse dimension on the large $L$ 
extrapolation. 
 Carrier density $n=1$. The conductivity obtained by 
extrapolating the $6 \times 6 \times L$ results are shown
as circles on the $y$ axes.
}
\end{figure}
low
frequency average ${\bar \sigma}_{av}(L)$, at some size $L$, as
the bulk d.c conductivity? Fig.6 illuatrates the extrapolation based on the
sequence $\{ L : 24, 32, 40, 48, 56, 64 \}$, at $\mu =0$, moving from weak
to strong disorder.

In the weak disorder
regime, the optical conductivity is `flat' for $\omega \ll \tau^{-1}$
so if low enough frequencies can be accessed (given the finite
size gaps), the d.c conductivity can be reasonably approximated.
This is the feature observed at $\Delta=2$ in Fig.6. However, in
the WL region and beyond, $\sigma(\omega)$ has non trivial frequency
dependence at low $\omega$, as evident in Fig.5.(b).
The corresponding 
low frequency average has significant $L$ dependence. 
Since $\sigma(\omega) \sim \sigma(0) + {\cal O}(\sqrt{\omega})$,
the low frequency average $\sigma_{av}(L) \sim \sigma(\infty) + 
{\cal O}(1/\sqrt L)$. The data at $\Delta =12, 16$ show a reasonable
fit to the 
square root form. 
The much stronger frequency dependence in the strong disorder regime
makes a size dependent study imperative.
We provide a discussion of the extrapolation scheme in an appendix.

These results illustrate the work involved in accessing the d.c
conductivity, particularly in the regime of strong disorder, where
a small $L$ calculation (at $L = 16$ say) might overestimate the
conductivity by a factor of $4$. This discrepancy worsens as
$\Delta \rightarrow \Delta_c$ and a systematic study of size
dependence is vitally important.

\subsection{Effect of the transverse dimension}

All the results quoted till now have been obtained via
extrapolation on a $6 \times 6 \times L$ geometry. The $6^2$
cross section was chosen to allow large $L$ to be accessed. 
However, it is important to quantify the error involved in 
chosing a specific transverse dimension $L_T$. To this end
we studied the low frequency average $\sigma_{av}$
in a sequence $L_T \times L_T \times L$ with $L_T =
 2, 4, 6, 8, 10, 12$ and $L = 10, 20, 30$, for
$\Delta = 2, 10$ and $16$, and $n=1$.
The averaging interval $\Delta \omega$ was scaled as 
$1/(L^2L_T)$ in all geometries.

Fig.7 shows $\sigma_{av}$ with respect to $L^{-1}$ for
the sequence $L_T$ specified above.
Panel $(a)$. shows the weak disorder, $\Delta = 2$ result.
Beyond $L_T = 4$ all the curves seem to converge to 
$\sigma \sim
0.70$ for $L \rightarrow \infty$. The extrapolation from 
$L_T = 6$, obtained using $L$ upto 64, is shown as a circle
on the $y$ axis, and is $\sim 0.68$.

For the strong disorder case, Fig.7.(b), the extrapolation for 
 $L_T =6$ is shown to be $\sim 0.001$,
while the  large $L_T$ data, using $L$ upto $30$,
 suggests that
the asymptotic value could be larger, $\sim 0.002$. 
This suggests that `small' $L_T$ somewhat underestimates
the conductivity (remember $L_T =1$ is one dimensional, 
so completely localised), while finite  $L$ overestimates
the conductivity. Except very close to the MIT these errors are
small for the sizes we use and, as verified by the phase diagram,
even the critical point is located to within $5 \%$.

\section{Scattering from magnetic disorder}

\subsection{Global features}

The effect of weak magnetic scattering on transport is quite
similar to that of potential scattering. The effect is 
contained in the Born scattering rate, $\tau^{-1}_s
\propto  N(\epsilon_F) J'^2 S^2$, and the weak 
coupling  resistivity $\rho(J', n) $ varies as $\sim b_1(n) J'^2$,
where $b_1(n)$ is a density dependent coefficient.
However, even at moderate coupling, $J' \sim 2$, new 
effects begin to show up in $\sigma(\mu)$. The conductivity
at half-filling, $n \sim 1$, gets suppressed 
more quickly than would be
guessed based on the Born argument.  This deviation, and its
evolution with increasing $J'$, arises from a fundamental
difference between potential scattering and magnetic scattering
on a `Kondo lattice'.

There are in fact {\it two} main   differences that show up
beyond weak coupling.
These are 
visible when we compare Figs.8-10, with Figs.1-4.
$(i)$~The conductivity in the potential scattering case decreases
monotonically (at fixed $n$) with increase in disorder, vanishing at
$\Delta_c(n)$, while in 
the magnetic scattering case,
at a {\it generic} density,  
the conductivity is {\it finite} even as $J' \rightarrow
\infty$. The resistivity `saturates' and there 
is no metal-insulator transition with increasing
$J'$, except in a narrow density window.
$(ii)$~The band center, $n \sim 1$, is of no particular
significance in the Anderson problem, except $\Delta_c$ being
largest. In the $J'$ problem the   response 
for $n \sim 1$ is dramatically different
from that in the rest of the band.
There is an MIT at  $J' \approx 5$.
These  differences
can be understood from an analysis of the
strong coupling end.

For $J'/t \gg 1$ it is useful to choose a local quantisation
axis at each site, for the electrons, parallel to the orientation
of the spin ${\bf S}_i$. The coupling $J'{\bf S}_i$ acts
 as a strong local
Zeeman field on the electron. Suppose the hopping term were absent.
The two local eigenfunctions
at each site would have spin projections parallel and antiparallel
to $J' {\bf S}_i$, with energy
$\mp J'/2$ respectively.
The zero  hopping problem leads to $N$ fold degenerate
levels at $\pm J'/2$. The `gap' $J'$ plays a key role at strong
coupling. The presence of hopping generates a degenerate perturbation 
on the locally aligned states (say), and the electrons can now
`hop' with an amplitude that depends on the orientation of
nearest neighbour spins. This mechanism has been extensively
discussed in the context of the double exchange model. 

The mixing introduced by `hopping' broadens the two levels into
bands.
For $J' \gtrsim W/2$ the broadening due to $t$ does not fill the
gap, {\it and the system is insulating at $n=1$}. For $J'$ 
below this critical value, $J_c$, say, the DOS at band center is
suppressed but finite, and the resistivity is still very large. 
In summary, the strong coupling physics of (incipient) band
splitting controls the resistivity close to band center, and 
creates an essential difference, in terms of $J'$ and $n$, 
with respect to standard Anderson localisation.

\begin{figure}
\centerline{
\psfig{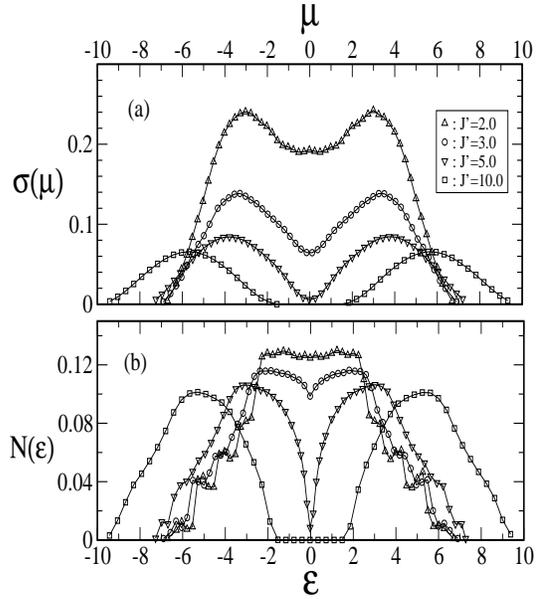}}
\vspace{.1cm}
\caption{ $(a)$. Conductivity as a function of Fermi energy, $(b)$.
 density of states, for different values of $J'$, in the case of
pure magnetic scattering.
}
\end{figure}

\begin{figure}
\vspace{.5cm}
\centerline{
\psfig{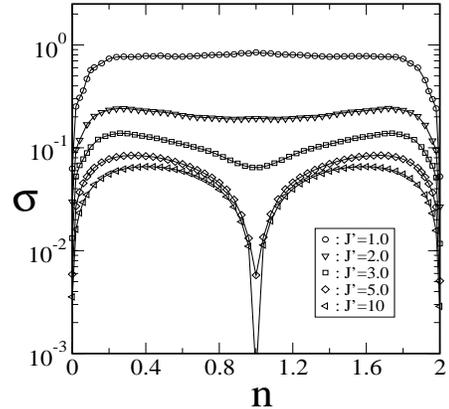}}
\vspace{.1cm}
\caption{ Dependence of conductivity on carrier density for 
varying $J'$. The conductivity scale is logarithmic.}
\end{figure}

\begin{figure}
\centerline{
\psfig{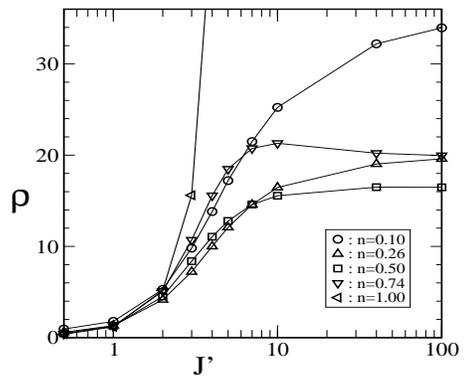}}
\vspace{.1cm}
\caption{ Resistivity variation with $J'$, for several electron density,
from the perturbative limit
to double exchange}
\end{figure}

The  saturation in $\rho(J')$ with increasing $J'$,
over most of the band, occurs because the effect of
large $J'$ is absorbed 
mainly in the band splitting. The effective
{\it disorder} seen by the electrons
comes from fluctuations in the hopping amplitudes, explained
in the next section, and these are ${\cal O}(t)$. 
The ratio
of fluctuation to mean hopping is moderate, so the large $J'$ limit
leads to a `dirty metal' but no metal-insulator transition.
This is 
unlike the Anderson problem where the electrons scatter off potential
fluctuations whose amplitude grows with increasing 
$\Delta$.

From the data in Figs.8-10 we can now 
identify the different
transport regimes.

\subsection{Transport regimes}

\subsubsection{Weak coupling: $J'/W \ll  1$}

The magnetic scattering rate $\Gamma_s$ is 
proportional to 
$  N(\epsilon_F)J'^2$,
and the weak coupling 
resistivity should be expandable in $\Gamma_s$.
The lowest order term is well known, corresponding to Born
scattering, with $\rho(J', n) \sim b_1(n) J'^2$. The density 
dependence is similar to that
 for potential scattering.
Assuming $\rho(J')$ to be analytic in $\Gamma_s$, {\it i.e}
ignoring possible log corrections etc, Fig.11 shows a fit
of the form $\rho(J', n) \sim b_1(n)J'^2 + b_2(n)J'^4$ 
to the low $J'$ resistivity.
The $J'^2$ character dominates upto $J' \sim 2$, as one can
see also in the $\sigma(n)$ plot in Fig.9, beyond which the
quartic term becomes important. We do not know if the coefficient 
of the quartic term has been analytically calculated, but the {\it sign}
of this term is crucial, and is 
density dependent, as we discuss  next.

\subsubsection{Intermediate coupling: $J'/W \sim {\cal O}(1)$ }

As is obvious from the  data in Fig.9-10, the resistivity saturates
with increasing $J'$, over most of the band. The 
\begin{figure}
\vspace{.4cm}
\centerline{
\psfig{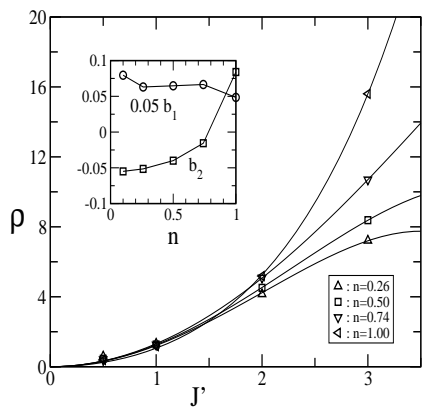}}
\vspace{.1cm}
\caption{ Fit to the weak coupling resistivity
of the form $\rho(J', n) \sim b_1(n)J'^2 + b_2(n)J'^4$. The symbols
are actual data and the firm lines are fits. The inset shows the 
$n$ dependence of the coefficients $b_1$ and $b_2$. Note the sign
change in $b_2$.} 
\end{figure}
exception is the
vicinity of $n=1$, and the lower edge of 
the band.
This suggests that the correction to the Born resistivity is 
{\it negative}
for $n$ away from $n=1$, and changes sign as $n \rightarrow 1$.
Fitting the data to $\rho(J', n) 
= b_1(n)J'^2 + b_2(n)J'^4$, the coefficient 
$b_2(n)$ illustrates the crossover from
saturation to escalation, as we move
across the band.  Fig.11 shows
the fit to this form and the coefficients are shown in the inset.
The `Born' coefficient is positive throughout the band, without
significant density dependence in the density interval shown.
The quartic coefficient changes sign, from positive
to negative, as $n$ is lowered from $1.0$ to $0.74$.

A confirmation of saturation or escalation cannot of course be
obtained 
from a low order expansion in $J'^2$, but even the `perturbative'
coefficient provides a hint of strong coupling physics.
It also suggests a smooth evolution from weak to strong coupling.

\subsubsection{Double exchange: $J'/W \rightarrow \infty$}

In the double exchange limit the $J'$ scale acts as a `constraint'
on the electron spin orientation and no longer directly affects
physical properties, the only effect is to renormalise the 
chemical potential. 
The mapping of the $J'/t \rightarrow \infty$ problem to a `spinless
fermion' problem with hopping dependent on nearest neighbour spin 
orientation has been widely discussed \cite{dex-map-ref}.
Transformed to spinless fermions, which correspond to 
original electron states with spin projection `locked'
parellel to the local quantisation axis, ${\bf S}_i$,
 the Hamiltonian becomes:
\begin{equation}
H = 
\sum_{\langle ij \rangle} t_{ij} \gamma^{\dagger}_i \gamma_j = 
\sum_{\langle ij \rangle} {\bar t}  \gamma^{\dagger}_i \gamma_j +
\sum_{\langle ij \rangle} \delta t_{ij} \gamma^{\dagger}_i \gamma_j 
\end{equation}
The  $t_{ij} $ being the spin orientation dependent hopping amplitude
specified earlier.
 We can split it into the mean (uniform)
hopping amplitude, ${\bar t}$, and the fluctuation $\delta t_{ij}$.

In the `extreme' 
paramagnetic phase of this model, the distribution of
hopping integrals is exactly known.  The spins are independently
distributed on a sphere so the $t_{ij}$ can be worked out.
There is no obvious small parameter, since both the mean value
of hopping, ${\bar t}$, as well as the fluctuation,
$\Delta t = \sqrt {\langle  {\delta t}^2 \rangle }$, are $\propto t$.
However, the ratio ${\Delta t}/{\bar t} \approx 1/3$. Numerical work
by Li {\it et al.} \cite{dex-loc}
 had demonstrated that less than $0.3 \%$ of 
states in the band are localised under this condition. It was
not clear whether the resistivity at the band center, $n=0.5$,
could be described within a Boltzmann approach. Narimanov and 
Varma \cite{narim-varm}
have demonstrated that the mean free path emerging from
the Boltzmann calculation is $l/a_0 \gtrsim 8$ so the method
is self-consistent. 

It seems now that
despite the localisation effects as $n \rightarrow 1$ and
$n \rightarrow 0$, resistivity over much of the band can
be understood within a effective `weak coupling' approach.
The resistivity is  $\approx  (0.1-0.2) \rho_{Mott} $ at the
band center according to our calculation.
The resistivity is also `particle-hole' symmetric, now 
within the lower band, but notice that this is
cleanly visible only at very large $J'/t$.

\subsubsection{Virtual orbital mixing: large finite $J'$}

As we move to finite $J'$ from the double exchange limit, the 
two bands still remain split (down to $J'/t \approx 5$) but
there is a virtual admixture that is introduced.
To access properties in this regime we need to use a
two orbital formulation, with the orbital energies still
separated by a large gap $\sim J'$. The chemical
potential remains in the lower band.
The two orbital model, written in terms of electronic states
with local quantisation axis, has the form:
\begin{equation}
H = \sum_{ij} t^{\alpha \beta}_{ij} \gamma^{\dagger}_{i \alpha} 
\gamma_{j \beta}
- \mu \sum_i n_i - {J' \over 2}
\sum_i (n_{i \alpha} - n_{i \beta})
\end{equation}

We have not seen a Boltzmann calculation of transport in
this regime, but using the two orbital formulation it might
be possible to set up such a scheme. The resistivity 
{\it decreases} as we move down from large $J'$, so
using the correct `basis' the transport may be accessible
within a Boltzmann approach (since the double exchange limit
is itself so accessible).
The increase in conductivity, $\delta \sigma(J')$, as we 
move to lower $J'$, is found to be proportional 
to $ 1/J'$.  
A perturbative correction to the large
$J'$ result, within a diagrammatic scheme yields the
same answer.

\subsubsection{Behaviour near band tails}

Spin disorder by itself cannot localise states in the 
center of the band, since ${\Delta t}/t$ is not large enough.
However at the band tails, {\it i.e},  $n$ close to $1$ or $0$,
the kinetic energy is small and a small fraction of states
can still be localised. As we have indicated earlier, this is
$\lesssim 0.3 \%$ 
\cite{dex-loc}
of the total number of states for $J' \rightarrow \infty$.

\begin{figure}
\vspace{.8cm}
\centerline{
\psfig{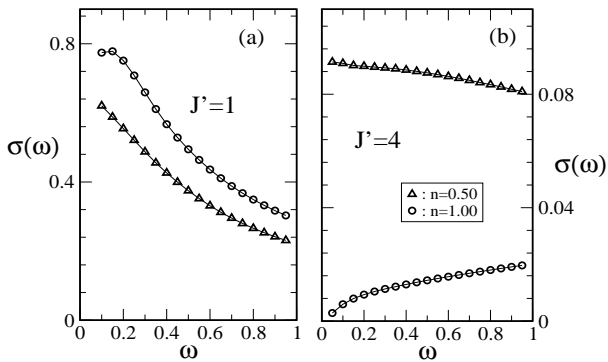}}
\vspace{.2cm}
\caption{Optical conductivity: $(a).$~Drude response at both $n=0.5$ and
$n=1.0$ at weak coupling, and $(b).$~strong scattering at $n=0.5$ and
alomst `insulating' response at $n=1.0$ at strong coupling.
}
\end{figure}

We do not know if any analytic approaches have been explored
in this localisation problem.
This regime would be relevant to the low doping magnetic
semiconductors, where there is also the possibility of
carriers trapping into spin polaronic states.

\subsection{Optical conductivity}

The optical conductivity confirms the trends seen in the
d.c conductivity. Fig.12 shows $\sigma(\omega)$ at $J'=1$
and $J'=4$, weak and `strong' coupling respectively,
at the center $(n=0.5)$ and edge $(n=1)$ of the lower 
band.

At weak coupling, over
the frequency range shown, $\sigma(\omega)$ is larger
at $n=1$, the center of the {\it full band}, compared to $n=0.5$.
The scattering rate is $\Gamma_s = 2 \pi N(\epsilon_F) J'^2$
which at the band center is $\approx 0.75$. If $\sigma(\omega)$
follows the Drude form then $\sigma(\Gamma_s)/\sigma(0)$ should
be $\sim 0.5$,  
which is consistent with Fig.12.(a). 
By $J'=4$, the trend has reversed. The $n=1$ case is almost insulating,
with $\sigma(0) \rightarrow 0$, while the conductivity at the (lower)
band center is finite and essentially flat on the scale considered.
This trend gets amplified as we go to even larger~$J'$.

\section{Combined structural and magnetic disorder}

\subsection{Global features}

In the presence of both structural and magnetic disorder it is not
possible to show the full density dependence of transport 
properties compactly, so we provide
two generic `cross sections' in Fig.13 at $n = 0.26$ and $n=1.00$. 
In addition to the effects  already noted for 
potential scattering and magnetic scattering,
 there are several novel features that arise.

$(a).$~For weak $J'$ and moderately large $\Delta$, 
{\it magnetic scattering weakens localisation effects}, 
as evident from
the intermediate $\Delta$ small $J'$ data in Fig.13.

$(b).$~At even
larger $\Delta$, where the system would have been 
Anderson localised, {\it magnetic
scattering converts the insulator to a metal}.
The critical disorder
$\Delta_c(n)$ shifts to a larger value $\Delta_c(n,J')$, see 
phase diagrams in Fig.14.

$(c).$~In contrast to purely magnetic scattering, where 
the resistivity typically `saturates' with increasing $J'$,
in the presence of structural disorder the system {\it can
go insulating with increasing $J'$}.

$(d).$~The `additivity' of magnetic and structural scattering
holds only over a very limited range in $\Delta$ and $J'$,
{\it 
Mathiessens rule generally does not hold}.

The major features, above,
 can be easily motivated after we 
write down the different {\it effective models} of scattering 
in the various transport regimes in the problem. 
Some of this has been discussed earlier by us \cite{sk-pm-epl},
so we will 
discuss mainly those aspects of the 
problem which 
have not been covered  earlier.

\subsection{Transport regimes}

The parameter space of the problem is large,
involving $n-\Delta-J'$, and it is convenient to first 
identify distinct density ranges and then classify the 
transport/scattering mechanisms.
The roughly distinct density regimes are the following.

(1).~The wide `mid band'
region $ 0.1  \lesssim n \lesssim 0.9$, of which $n=0.26$ in
Fig.13.$(a)$ is typical, and we have discussed the $n=0.5$ case
earlier \cite{sk-pm-epl}. 

(2).~$n\rightarrow 1$, where the
response is similar to the mid band region at weak $J'$, but 
the large $J'$ response is distinctive, Fig.13.$(b)$.

(3).~$n \rightarrow 0$, where structural disorder and magnetic
scattering readily leads to localisation.

\subsubsection{Generic density: the mid band region}

Let us consider this typical density regime first.
For generic densities, $0.1 \lesssim n \lesssim 0.9$, say,
there are tentatively five different 
transport regimes in the
problem.
These are:
$(i)$~Both   $\Delta$ and $J'$ small:
the weak scattering regime, where  the effect of structural
disorder and magnetic scattering are perturbative and additive.
$(ii)$~Moderate $\Delta$ and small $J'$:
spin flip correction to weak localisation.
The $\Delta$ dependence shows  WL
corrections and
spin flip scattering weakens the WL correction.
$(iii)$~Large $\Delta$, $\sim \Delta_c$, and small $J'$:
spin dephasing driven insulator-metal transition (IMT). 
$(iv)$~
$J'/t \rightarrow \infty $, with 
varying $\Delta$:
the disordered double exchange (DE) limit.
$(v)$~$J'/t \gg 1$ but finite, and moderate to large $\Delta$:
the intermediate coupling `metal'.

\begin{figure}
\vspace{1.0cm}
\centerline{
\psfig{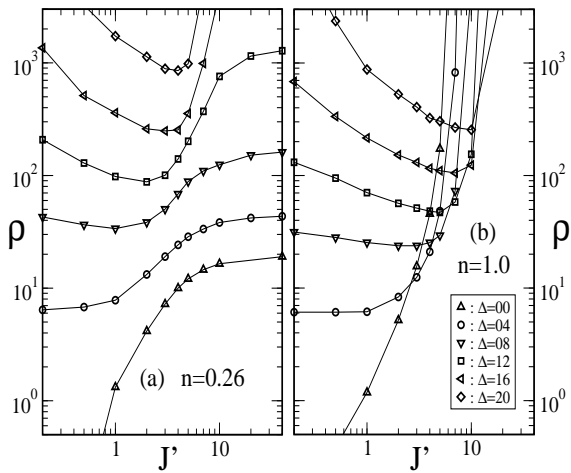}}
\vspace{.3cm}
\caption{Global behaviour of the resistivity with varying structural 
disorder $(\Delta)$ and magnetic coupling $(J')$: $(a).$~$n=0.26$ 
typical of most of band, and $(b).$~$n=1.0$ which, at large $J'$, 
corresponds to the upper edge of the lower band and has its own
distinct transport response.}
\end{figure}
\begin{figure}
\centerline{
\psfig{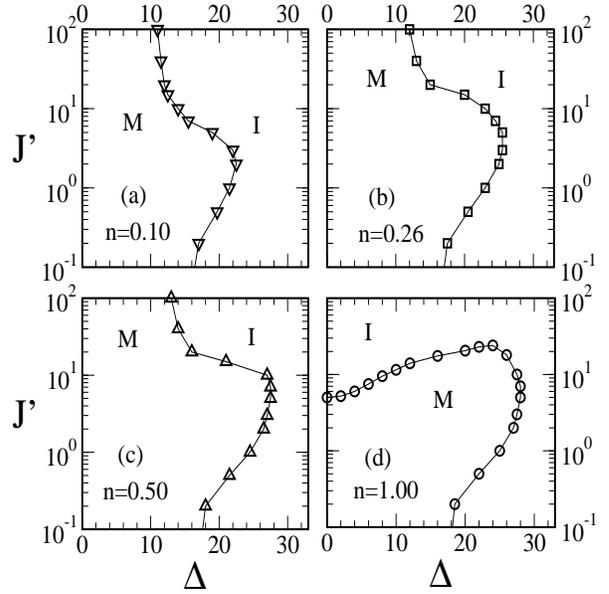}}
\vspace{.2cm}
\caption{ Putting together different constant density cross-sections to
create a `global' insulator-metal phase diagram for the $\Delta + J'$ 
problem. 
The densities
are marked in the panels. 
M stands for a metallic phase while I is insulating. 
The bounding curves can be viewed as $\Delta_c(n, J')$.
Notice the log scale on $J'$.}
\vspace{.3cm}
\end{figure}

$(i)$~When $\Delta$ and $J'$ are both small the transport
can be understood in terms of
additive Born scattering,  with the net 
scattering rate, $\Gamma(\Delta, J') \approx a_1\Delta^2 + b_1 J'^2 $,
and the resistivity $\rho \propto \Gamma (\Delta, J')$.
The `window' describing this regime is 
roughly  $J' \le 3$ and $\Delta \lesssim 4$.
The resistivity in this regime is  $\rho  <  0.1 
\rho_{Mott}$,
{\it i.e}, below  $100~\mu \Omega$cm, say.
This corresponds to the bottom left hand corner in Fig.13.(a) and,
as our earlier data showed \cite{sk-pm-epl}, Mathiessens rule holds.

$(ii)$~At larger $\Delta$ remaining at small $J'$,
as the WL  
corrections show up, 
spin flip scattering 
\cite{and-loc-sf1,and-loc-sf2} 
of the electrons 
by the random magnetic moments reduces the  localising effect
of structural disorder, {\it i.e}, 
$\partial \rho/\partial J'\vert_{n, \Delta}  < 0$.
Just as inelastic scattering weakens quantum interference by 
introducing decoherence, spin flip
scattering leads to {\it spin decoherence}. We have
quantified the $\Delta$ and $J'$ dependence of the
effect in the earlier paper \cite{sk-pm-epl}.

$(iii)$~At even larger disorder, $\Delta \gtrsim \Delta_c$,
where the $J'=0$ system would have been
Anderson localised, spin flip scattering opens up a metallic
window. The structural disorder needed for localisation shifts to
a larger value,  {\it i.e}, $\partial \Delta_c/\partial J' \vert_{n} >0$.
This effect in visible in all the panels in Fig.14.

$(iv)$~Now consider the DE limit, $J' \rightarrow \infty$.
As we have discussed in  Section IV, the form of the 
resistivity $\rho(J', \Delta=0)$ arising from `magnetic disorder' 
at large
$J'$ is very different from what one observes in $\rho(J'=0, \Delta)$
at large $\Delta$. This is because $J'$ contributes to both
`band splitting' and effective disorder, and the effective
disorder saturates as $J'/W \rightarrow \infty$ with 
$J'$ controlling only the band splitting.
The presence of structural disorder in the $J' \rightarrow \infty$ problem
strongly enhances the resistivity and localising tendency.
Using the transformations used in the previous section:
\begin{eqnarray}
H& =& \sum_{ij} t_{ij} (\theta, \phi) 
\gamma^{\dagger}_i \gamma^{~}_j +
\sum_i \epsilon_i \gamma^{\dagger}_i \gamma^{~}_i \cr
& \equiv&
\sum_{ij} t_0 
\gamma^{\dagger}_i \gamma_j +
\sum_{ij} \delta t_{ij} \gamma^{\dagger}_i \gamma_j +
\sum_i \epsilon_i \gamma^{\dagger}_i \gamma_i
\end{eqnarray}
The localisation properties
of this model have been studied by Li {\it et al.} \cite{dex-loc}, 
although 
they did not calculate the resistivity.
The `hopping disorder' by itself localises less than
$0.5 \%$ of the states in the band.
On adding structural disorder
the mobility edge moves inward, with localisation of the
full band  occuring at 
$\Delta/t \sim 11.5$, which can be approximately understood from
the roughly $30 \%$ band narrowing due to spin disorder.

$(v)$~Finally 
the regime with large but finite $J'$ and strong structural disorder.
We have seen that the effect of {\it  small $J'$} at strong
structural disorder  
can be qualitatively understood in terms of
the spin dephasing effect on Anderson localisation. However,
the small $J'$ behaviour with $\partial \rho/\partial J' < 0$ 
quickly leads to a minimum and 
then a regime with $\partial \rho/\partial J' > 0$. 
Such behaviour can be viewed as an extension of
the $J'^2$ term seen at weak disorder, but it is more fruitful to
approach the effect 
from the strong coupling
DE  end, as we do below. 
Transforming to the usual local spin quantisation frame and  retaining {\it both}
 the parallel and anti-parallel electron states, we have:
\begin{equation}
 H =  \sum_{ij} t^{\alpha \beta}_{ij} \gamma^{\dagger}_{i \alpha} 
\gamma_{j \beta}
+ \sum_i \epsilon_i n_i 
- {J' \over 2} \sum_i (n_{i \alpha} - n_{i \beta})
\end{equation}
The major source of disorder is still $\epsilon_i$, with
additional contribution from the $\delta t_{ij}^{\alpha \beta}$.
The orbital mixing effect of 
`off diagonal' couplings, either in terms of mean amplitude
or fluctuations, is regulated  by the large energy denominator
$J'$.
Although the `reference' problem, $J' \rightarrow \infty$, is
not analytically tractable in the presence of structural disorder,
it can be shown that  orbital mixing
generates a  correction to conductivity $ \sim {\cal O}(1/J')$.

\subsubsection{Half-filling: $n \rightarrow 1$}

For $n \rightarrow 1$, the effects at small $J'$ are 
similar to $(i)-(iii)$  at generic densities, discussed above.
This is borne out by the behaviour of $\rho(J', \Delta)$ in Fig.13.(b)
and the phase diagram in Fig.14.(d). 
At large $J'$, however, 
the system always goes insulating, see Fig.13.(b), as we have
discussed in Section IV as well. This effect is obviously due to 
the band splitting induced by large $J'$ and the vanishing DOS,
$N(\epsilon_F)$, at $n=1.0$. Thus, for $n=1.0$, the metallic
phase is bounded both in $\Delta$ and $J'$, Fig.14.(d).

There is however an interesting and possibly unexpected feature in
Fig.14.(d) for $J \gtrsim 5.0$, where the $\Delta=0$ system 
becomes insulating. We may have imagined that introducing structural
disorder in this system would enhance localisation. This 
however is not true, and structural disorder actually `metallises' the
reference band split state, and the critical $J'$ needed for localisation
{\it increases in the presence of structural disorder.
}

The origin of the effect above lies in the `band broadening' effect of
structural disorder. The $\Delta=0$ problem had a narrow (vanishingly
small) gap in the DOS, and the presence of structural disorder 
creates finite DOS at
the Fermi level, effectively closing the gap. Since the net  disorder
arising from the random spins and the structural disorder is relatively
weak the finite DOS seems sufficient to lead to a metallic, albeit
highly resistive, phase. For $\Delta \sim 4$ and $J' \sim 5$, the 
resistivity is roughly $0.5 \rho_{Mott}$. As $\Delta$ becomes large, or
$J'$ becomes large, this metallic window is lost due to the effects either
of $\Delta$ driven localisation or band splitting.

\subsubsection{Very low density: $n \rightarrow 0$}

The case of $n \rightarrow 0$, for example  $n=0.01$, say, is
unfortunately hard to access  with control
for the system sizes that we have used. We expect that the small
$J'$ behaviour will be similar to that in the rest of the band, with 
enhanced resistivity (due to the low carrier density) 
while the  behaviour for $J' \rightarrow \infty$
will be
similar to that for $n \rightarrow 1$ (due to the particle-hole 
symmetry within the lower band, $0 < n < 1$).
Localisation in the $n \rightarrow 0$ limit, 
we believe, is better explored via transfer matrix methods
due to the large accessible size.

\section{Concluding remarks}

In this paper we have presented controlled results on electron 
transport in the background of arbitrary structural and spin disorder
and provided a framework within which the data can be
analysed. We benchmarked our Kubo formula based 
method in the standard problem 
of potential scattering and Anderson localisation.
We then explored 
the distinct transport regimes that arise in 
the case of pure magnetic scattering, as well as 
the combined
effect of structural and magnetic disorder. 
In contrast to the effect of only structural disorder 
(where the resistivity
`escalates' with increasing disorder) or 
only magnetic scattering (where it
`saturates' with increasing disorder) their  combined action can lead to
non monotonic dependence and novel transport regimes.
The method developed in this paper can be directly taken
over in calculating the resistivity in the presence of annealed disorder,
where accessible system sizes rarely exceed $\sim 10^3$, 
and has been extensively used by us in Monte Carlo studies of
several ``disordered'' electron~systems.

\vspace{.2cm}

We acknowledge use of the Beowulf cluster at HRI.
S.K. gratefully acknowledges support by the Deutsche
Forschungsgemeinschaft through SFB484.

\section{Appendix: Extrapolation for the d.c conductivity}

It is known that 
at sufficiently low frequency the optical
conductivity in the 3D Anderson model 
follows a simple power law \cite{and-loc2}
with the exponent depending
on the strength of disorder and electron density. This means
that we can write: $ \sigma(\omega) = A + B \omega^{\alpha}$,
where 
the coefficient $A \ge 0$ is $\sigma_{dc}$, 
while the next term gives the leading
low frequency correction. 
This form captures all the broad regimes in 3D. For example, 
at half-filling:
$(i)$~low disorder, Born scattering, $\Delta \ll \Delta_c$, gives
 $B <0$, $\alpha=2$, $(ii)$~moderate disorder, weak localisation 
corrections: $B >0$, $\alpha =1/2$, $(iii)$~critical disorder, 
$\Delta = \Delta_c$: $A=0$, $B >0$, $\alpha=1/3$, and $(iv)$~localised 
phase, $\Delta > \Delta_c$: $A=0$, $B >0$, $\alpha=2$.

The form for $\sigma(\omega)$ fixes the form for $\sigma_{av}
(\Delta \omega, L)$. Setting $\Delta \omega \sim 1/L$, and using
the form for $\sigma(\omega)$ above,  we obtain the
three parameter form for $\sigma_{av}(L)$: $\sigma_{av}(L) 
\sim  A +  {B \over {(\alpha  + 1)}} L^{-\alpha} $.
The extrapolation is a least square 
three parameter fit to our $L$ dependent
data, and has enough flexibility to cover
{\it all disorder regimes}.

In practice, a simpler
two parameter extrapolation also works reasonably as long 
as one is in the metallic phase, even close to the metal-insulator
transition: 
$\sigma_{av}(L) \sim A +  {B \over (3/2)} L^{-1/2}
$ 
This
derives from $\sigma(\omega) \sim A + B \omega^{1/2}$.
Using this restricted fitting function, the MIT 
can be roughly located when $A$, the d.c conductivity, falls
below a preset limit, $10^{-6}$ say, (which in absolute units
is a resistivity $\sim 10^4 \rho_{Mott}$). Having located the
transition approximately, the more elaborate three parameter
fit can be used to confirm the metallic/insulating character on
two sides of the critical point.
Our MI phase diagrams are constructed using this strategy.

{}

\end{document}